# Multi-Agent Routing and Scheduling through Coalition Formation

Luca Capezzuto
Electronics and Computer Science
University of Southampton
Southampton, UK
luca.capezzuto@soton.ac.uk

Danesh Tarapore
Electronics and Computer Science
University of Southampton
Southampton, UK
d.s.tarapore@soton.ac.uk

Sarvapali D. Ramchurn
Electronics and Computer Science
University of Southampton
Southampton, UK
sdr1@soton.ac.uk

## ABSTRACT

In task allocation for real-time domains, such as disaster response, a limited number of agents is deployed across a large area to carry out numerous tasks, each with its prerequisites, profit, time window and workload. To maximize profits while minimizing time penalties, agents need to cooperate by forming, disbanding and reforming coalitions. In this paper, we name this problem *Multi-Agent Routing and Scheduling through Coalition formation* (MARSC) and show that it generalizes the important Team Orienteering Problem with Time Windows. We propose a binary integer program and an anytime and scalable heuristic to solve it. Using public London Fire Brigade records, we create a dataset with 347588 tasks and a test framework that simulates the mobilization of firefighters. In problems with up to 150 agents and 3000 tasks, our heuristic finds solutions up to 3.25 times better than the Earliest Deadline First approach commonly used in real-time systems. Our results constitute the first large-scale benchmark for the MARSC problem.

## 1 INTRODUCTION

Assignment problems form a core part of combinatorial optimization [2] that boasts successful applications in many real-world domains, such as transportation management, production planning, supply chain, market clearing, and so on [17]. For this reason, these problems have also been extensively studied from the perspective of multi-agent and multi-robot systems [21]. A main class of multi-agent assignment problems is *cooperative coordination*, which considers the synergies between agents that have common goals [25]. In multi-robot systems, it is called Multi-Robot Task Allocation [5]. Within this class, we are interested in problems that focus both on routing and scheduling, which have gained importance in the last years [15].

We are motivated by the scenarios considered in disaster robotics [14], especially the RoboCup Rescue League [9]. Natural and man-made disasters, such as the Australian bushfire season or the Beirut explosion in 2020, lead to severe loss of life and significant damage to infrastructure. Responding to these events includes complex actions such as extinguishing fires, clearing the streets and evacuating civilians. If the number of first responders is limited, they need to cooperate to act as fast as possible, since any delay can lead to further tragedy and destruction [1]. Cooperation is also necessary when tasks require combined skills. For example, to extract survivors from the rubble of a collapsed building, rescue robots detect life signs with their sensors, firefighters dig and paramedics load the injured into ambulances. Depending on availability, ambulances can transport the injured to different hospitals, and survivors can be transferred to various evacuation centers. There may be task precedences, such as when firefighters have to clear a road to allow ambulances to enter an area. Some tasks may be more critical or urgent than others. For example, saving lives has a greater *profit* (i.e., it is more important) than clearing the rubble. At any moment, new fires could break out or other buildings could collapse, thus first responders must be ready to deploy to other areas.

In the problem we are considering, tasks are spatially distributed and have profits, precedences, time windows, workloads and multiple possible locations. To maximize profits while minimizing time penalties, the agents need to form, disband and reform coalitions. A *coalition* is a flat and short-lived organisation of agents that performs tasks more effectively or quickly than single agents [3]. By forming coalitions, the agents define in what order tasks are performed (routing) and at what times (scheduling).

Many similar problems have been studied to date [11, 15, 21]. We mention below those that come closest to our target. Scerri et al. [23] were the first to investigate time windows and interdependent simultaneous tasks, while Vig and Adams [28] applied the seminal work of Shehory and Kraus [24] to multi-robot systems. However, both works do not consider situations where task scheduling is required. Zlot [30] defined a problem where tasks are decomposable in multiple ways, but do not have precedence relations. Ramchurn et al. [20] proposed a coalition formation problem with spatial and temporal constraints, but without taking into account task precedences, time windows and multiple possible locations per task. Korsah [10] investigated spatio-temporal constraints, task dependencies and multiple possible locations per task, while Godoy and Gini [6] and Nunes et al. [16] studied problems with time windows, spatial and precedence constraints. Nonetheless, all three works ignore coalition formation.

To the best of our knowledge, no existing model captures our problem in its entirety. Against this background, we propose the following contributions:

(1) The *Multi-agent Routing and Scheduling through Coalition formation* (MARSC) problem, the first generalization of the *Team Orienteering Problem with Time Windows* (TOPTW) [27] that considers coalition formation and can be used in real-time domains such as disaster response.

(2) A mathematical programming formulation of the MARSC, and an anytime and scalable algorithm, both simple and easy to implement.
(3) The first large-scale MARSC benchmark, based on real-world data published by the London Fire Brigade [12, 13].

The rest of the paper is organized as follows. Section 2 formulates our problem, while Section 3 presents our algorithm. Section 4 evaluates our algorithm in our test framework and Section 5 concludes. Appendix A shows how the MARSC generalizes the TOPTW.

## 2 PROBLEM FORMULATION

We formulate the MARSC as a *Binary Integer Program* (BIP) [29]. We give our definitions, then detail our decision variables, constraints and objective function.

Since our problem generalizes the TOPTW, we prefer to use the term *node* instead of *task*, and the expression *visiting a node* instead of *performing a task*.

### 2.1 Definitions

Let $V = \{v_1, \ldots, v_m\}$ be a set of $m$ nodes and $A = \{a_1, \ldots, a_n\}$ be a set of $n$ agents. Let $L$ be the finite set of all possible node and agent locations. Time is denoted by $t \in \mathbb{N}$, starting at $t = 0$, and agents travel or visit nodes with a base time unit of 1. The time units needed by an agent to travel from one location to another are given by the function $\rho : A \times L \times L \to \mathbb{N}$. Having $A$ in the domain of $\rho$ allows to characterize different agent features (e.g., speed or type). Let $l_a^t \in L$ be the location of agent $a$ at time $t$, where $l_a^0$ is the initial location of $a$ and is known a priori.

*2.1.1 Node demand.* Each node $v$ has a *demand* $D_v = (L_v, w_v, \phi_v, \alpha_v, \beta_v, \gamma_v)$, where:

- $L_v \subseteq L$ is the *set of possible locations* of $v$ [10].
- $w_v \in \mathbb{R}_{\geq 0}$ is the *workload* of $v$, or the amount of work required to visit $v$ [3].
- $\phi_v \in \mathbb{R}_{\geq 0}$ is the *profit* of $v$, or the reward associated with the visit of $v$ [27].
- $[\alpha_v, \beta_v, \gamma_v]$ is the *time window* of $v$, such that $\alpha_v \in \mathbb{N}$ is the *earliest time* of $v$, or the time starting from which agents can visit $v$, $\beta_v \in \mathbb{N}$ is the *soft latest time* of $v$, or the time until which agents can visit $v$ without incurring in a penalty, and $\gamma_v \in \mathbb{N}$ is the *hard latest time* of $v$, or the time until which agents can visit $v$ incurring in a penalty [15].

We assume that $\alpha_v \leq \beta_v \leq \gamma_v, \forall v \in V$, and call $t_{max} = \max_{v \in V} \gamma_v$ the *maximum problem time*.

*2.1.2 Node order.* Let $Prec \subseteq V \times V$ be such that if $(v_1, v_2) \in Prec$ then $v_1$ must be visited before $v_2$. Each $(v_1, v_2) \in P$ is called a *precedence* and can be graphically denoted with $v_1 \to v_2$. We assume that $G = (V, Prec)$ is a finite, directed and acyclic graph. Moreover, without loss of generality, we assume that $Prec$ does not contain relations which can be inferred transitively, that is: $(v_1, v_2) \in Prec \land (v_2, v_3) \in Prec \Rightarrow (v_1, v_3) \notin Prec$.

*2.1.3 Coalition and coalition value.* A subset of agents $C \subseteq A$ is called a *coalition*. For each coalition, node and location there is a *coalition value*, given by the function $u : P(A) \times V \times L \to \mathbb{R}_{\geq 0}$, where $P(A)$ is the power set of $A$. The value of $u(C, v, l)$ is the amount of work that coalition $C$ does on node $v$ at location $l \in L_v$ in one time unit. In other words, when $C$ visits $v$ in $l$, $u(C, v, l)$ expresses how well the agents in $C$ work together, and the workload $w_v$ decreases by $u(C, v, l)$ at each time.

### 2.2 Decision variables

We use the following indicator variables:

$$\forall v \in V, \forall l \in L_v, \forall t \in [\alpha_v, \gamma_v], \forall C \subseteq A, \ x_{v,l,t,C} \in \{0,1\} \quad (1)$$

where $x_{v,l,t,C} = 1$ if node $v$ in location $l$ and at time $t$ is visited by coalition $C$, and 0 otherwise.

### 2.3 Constraints

There are 4 types of constraints: structural, temporal, spatial and ordering. We formalize each below.

*2.3.1 Structural constraints.* Each node can be visited by at most one coalition at each time:

$$\forall v \in V, \forall l \in L_v, \forall t \in [\alpha_v, \gamma_v], \ \sum_{C \subseteq A} x_{v,l,t,C} \leq 1 \quad (2)$$

*2.3.2 Temporal constraints.* Each node can be visited in at most one location, only within its time window, and no agent can work on it after its workload has been completed. This is achieved by introducing, for each node $v$ and location $l \in L_v$, an auxiliary binary variable $y_{v,l} \in \{0, 1\}$, and the following constraints:

$$\forall v \in V, \sum_{l \in L_v} y_{v,l} \leq 1 \quad (3)$$

$$\forall v \in V, \forall l \in L_v,$$
$$\sum_{t \in [\alpha_v, \gamma_v]} \sum_{C \subseteq A} \lceil u(C, v, l) \rceil \cdot x_{v,l,t,C} = \lceil w_v \rceil \cdot y_{v,l} \quad (4)$$

*2.3.3 Spatial constraints.* An agent cannot visit a node before reaching one of its possible locations. This identifies two cases: when an agent reaches a node location from its initial location, and when an agent moves from one node location to another. The first case imposes that, for each node $v$, location $l \in L_v$ and coalition $C$, the decision variable $x_{v,l,t,C}$ can be positive only if all agents in $C$ can reach $l$ at a time $t' < t$:

$$\forall v \in V, \forall l \in L_v, \forall C \subseteq A,$$
$$\text{if } \lambda = \max_{a \in C} \rho(a, l_a^0, l) \geq \alpha_v \text{ then } \sum_{t \in [\alpha_v, \lambda]} x_{v,l,t,C} = 0 \quad (5)$$

The value of $\lambda$ is the maximum time at which an agent $a \in C$ reaches $l$, from its initial location at time $t = 0$. Conditional constraints are usually formulated using auxiliary variables or the Big-M method. However, such approaches further enlarge the mathematical program or cause numerical issues [7, Section 5.4.2]. Consequently, in the preprocessing step necessary to create our BIP, we can implement Equation 5 simply by excluding the variables that must be equal to zero.

The second case requires that if an agent cannot visit two nodes consecutively, then it can visit at most one:

$$\forall v_1, v_2 \in V, \forall l_1 \in L_1, \forall l_2 \in L_2, \forall C_1, C_2 \subseteq A : C_1 \cap C_2 \neq \emptyset,$$
$$\forall t_1 \in [\alpha_1, \gamma_1], \forall t_2 \in [\alpha_2, \gamma_2] : t_1 + \max_{a \in C_1 \cap C_2} \rho(a, l_1, l_2) \geq t_2, \quad (6)$$
$$x_{v_1, l_1, t_1, C_1} + x_{v_2, l_2, t_2, C_2} \leq 1$$

Hence, coalition $C_2$ can visit node $v_2$ only if all agents in $C_1 \cap C_2$ can reach location $l_2$ within the hard latest time $\gamma_2$. Equation 6 also implies that two nodes cannot be visited by the same agent at the same time. Consequently, coalitions that exist in different locations at the same time are disjoint.

There are no synchronization constraints [15]. Thus, when a node $v$ in location $l$ is allocated to a coalition $C$, each agent $a \in C$ starts working on $v$ as soon as it reaches $l$, without waiting for the remaining agents. This means that $v$ is completed by a temporal sequence of subcoalitions of $C$: $\exists S \subseteq P(C)$ such that $\forall C' \in S$, $\exists t \leq \gamma_v, x_{v,l,t,C'} = 1$, where $P(C)$ is the power set of $C$.

*2.3.4 Ordering constraints.* If $v_1 \rightarrow v_2$ then $v_2$ can only be visited after $v_1$. This means that if $\alpha_2 \leq \gamma_1$ (i.e., the time windows overlap), then no coalition can work on $v_2$ at time $t \in [\alpha_2, \gamma_1]$:

$$\forall v_1 \in V \text{ if } \exists v_2 \in V : v_1 \rightarrow v_2 \wedge \alpha_2 \leq \gamma_1 \text{ then}$$
$$\forall l \in L_2, \forall C \subseteq A, \sum_{t \in [\alpha_2, \gamma_1]} x_{v_2, l, t, C} = 0 \quad (7)$$

No ordering constraints are required if $\alpha_2 > \gamma_1$. Similar to Equation 5, we can implement Equation 7 by excluding variables instead of penalizing them.

## 2.4 Objective function

Let $x$ be a *solution*, that is, a value assignment to all decision variables, which defines the route and schedule of each agent. For each node $v$ and time $t$, let

$$\psi_{v,t} = \begin{cases} 1, & \text{if } t \leq \beta_v \\ 1 - \frac{t - \beta_v}{\gamma_v - \beta_v + 1}, & \text{otherwise} \end{cases} \quad (8)$$

bet the *penalty* of visiting $v$ during $t$, and let

$$score(x) = \sum_{v \in V} \sum_{l \in L_v} \sum_{t \in [\alpha_v, \gamma_v]} \sum_{C \subseteq A} \phi_v \cdot \psi_{v,t} \cdot x_{v,l,t,C} \quad (9)$$

be the *score* of $x$, which expresses how much $x$ maximizes the total profit and the number of nodes that are visited within their soft latest times (thus minimizing the total time penalty). The objective of the MARSC is to find a solution with maximum score and such as to satisfy all constraints:

$$\arg\max_x score(x) \text{ subject to Equations 1 – 7} \quad (10)$$

Both creating all decision variables (Equation 1) and finding an optimal solution exhaustively (Equation 10) may require to list all $\mathcal{L}$-tuples over $P(A)$, where $\mathcal{L} = |V| \cdot |L| \cdot t_{max}$. This implies a worst-case time complexity of:

$$O\left(\left(2^{|A|}\right)^{\mathcal{L}}\right) = O\left(2^{|A| \cdot |V| \cdot |L| \cdot t_{max}}\right) \quad (11)$$

THEOREM 2.1. *The MARSC generalizes the TOPTW.*

The proof of Theorem 2.1 is given in Appendix A. Since the TOPTW is NP-hard [27], the MARSC is also NP-hard.

## 3 AN ANYTIME AND SCALABLE ALGORITHM

A trivial way to solve the MARSC would be to implement Equation 10 with solvers such as CPLEX or GLPK. Although this would guarantee anytime and optimal solutions, it would also take exponential time to both create and solve the BIP (Equation 11). This limits this practice to offline contexts or very small problems. For example, using CPLEX 20.1 with commodity hardware, we can solve within hours problems with superadditive coalition values (i.e., $u(C, v, l) = |C|$), uniformly distributed workloads and time windows, locations based on the taxicab metric, and $dim = |A| \cdot |V| \cdot |L| \leq 25$. With greater $dim$ values, the runtime increases rapidly to days.

For this reason, this section presents an anytime and heuristic algorithm that trades optimality for scalability, called *Bounded Node Traversal* (BNT). We begin by explaining its solution method, then present and analyze the algorithm.

### 3.1 Solution method

Let $P(V)$ be the power set of $V$ and $P_i(V) \subseteq P(V)$ be such that each $\varphi \in P_i(V)$ contains $i$ nodes. Let $x_\varphi$ denote a solution to the nodes in $\varphi$ that satisfy all constraints. We can solve the MARSC as follows:

(1) Starting with $i = 1$, find a solution $x_\varphi, \forall \varphi \in P_i(V)$.
(2) Compute $\varphi_i^* = \arg\max_{\varphi \in P_i(V)} score(x_\varphi)$.
(3) Define the *expansion* of $\varphi_i^*$:

$$P_{i+1}^*(V) = \{\varphi_i^* \cup \{v\} : v \in V \setminus \varphi_i^*\}$$

Hence, each $\varphi \in P_{i+1}^*(V)$ contains $\varphi_i^*$ plus a new node.
(4) Set $P_{i+1}(V) = P_{i+1}^*(V)$ and $i \leftarrow i + 1$.
(5) Repeat Steps 1 – 4 until $i > |V|$ or $\nexists \varphi \in P_i(V)$ such that $\forall \varphi' \in P_{i-1}(V), score(\varphi) > score(\varphi')$.

At each step, at most $i$ nodes are visited and at most $|V| - i$ nodes must be visited. Hence, the number of node traversals in this best-first search [4] is:

$$|V| + (|V| - 1) + \cdots + 1 = \sum_{i=0}^{|V|} (|V| - i) \in O(|V|^2) \quad (12)$$

### 3.2 Algorithm design and analysis

Algorithm 1 implements the method of Section 3.1. Let $x_i \subseteq x$ denote a *singleton solution*, that is, a solution to node $i$. Since it makes sense to find solutions to nodes with the earliest time windows first, the sorting at line 2 allows the loop at line 5 to define $x$ as a schedule of singleton solutions. The agents decide to form, disband and reform coalitions as follows. For each node $v$ not yet visited and location $l \in L_v$, lines 6 – 7 choose a coalition $C$ such that:

- Each agent $a \in C$ is not reaching nor visiting another node and is closest to $l$ than to the location of any other node not yet visited.
- $C$ has minimum size and can visit $v$ within $\gamma_v$.

At line 7, to define $C$, all the available agents are sorted by arrival time to $l$. This coalition formation technique is guaranteed to converge to a solution that satisfies the spatio-temporal constraints [3, Theorem 1]. At line 8, since $score(\cdot)$ is an additive function (Equation 9), we maximize the total solution score by maximizing the score of each singleton solution.

Algorithm 1 is anytime since if the loop at line 3 was stopped prematurely, it would return a solution to the first $i - 1$ nodes. If it cannot find a solution to node $v_i$, then the same holds for each subsequent node in $V_\alpha$. In this case, line 10 stops the whole procedure. Sorting $V$ at line 2 requires $\Omega(|V| \log |V|)$ time [4]. Likewise, sorting $A$ at line 7 requires $\Omega(|A| \log |A|)$ time. Assuming $O(|A|)$

**Algorithm 1:** *Bounded Node Traversal* (BNT)

**Input:** nodes $V$, demands $\{D_v\}_{v \in V}$, order $P$, agents $A$

1  $x \leftarrow$ empty vector // the solution to return
2  $V_\alpha \leftarrow$ sort $V$ by earliest time, while satisfying the ordering constraints
3  **for** $i \leftarrow 1; i \leq |V|; i \leftarrow i+1$ **do**
4      $x_i \leftarrow$ NIL // solution for node $i$
5      **for** $v \in V_\alpha$ and $l \in L_v$ **do**
6          $A_{v,l} \leftarrow$ agents that can reach $l_v$ within $\gamma_v$
        // satisfy the spatial constraints
7          $x_{v,l} \leftarrow$ define from $A_{v,l}$ a solution where the coalition has minimal size and is the closest to $v$
        // satisfy the temporal constraints
8          **if** $score(x_{v,l}) > score(x_i)$ **then**
9              $x_i \leftarrow x_{v,l}$ // maximise score (Eq. 10)
10     **if** $x_i =$ NIL **then** // $\nexists$ singleton solution $x_{j \geq i}$
11         **break**
12     Update the allocation history of agents // satisfy the structural constraints
13     Remove from $V_\alpha$ the node visited by $x_i$
14 **return** $x$

and $O(1)$ time for lines 12 and 13, respectively, and considering Equation 12, the overall time complexity is:

$$O\left(|V|^2 \cdot |L| \cdot |A| \log |A|\right) \quad (13)$$

If we execute Algorithm 1 again, avoiding the traversals already made, we may obtain a better solution. In particular, we can find a solution to each $\varphi \in P(V)$ in $O(|V|!)$ runs. This implies that each execution establishes a lower bound for the number of visitable nodes, and the more we re-execute, the more the bound improves.

## 4 EMPIRICAL EVALUATION

We created a dataset[1] with 347588 nodes using open records published by the London Fire Brigade over a period of 11 years. Then, we wrote a test framework in Java[2] and compared BNT against our implementation of the *Earliest Deadline First* (EDF) scheduling commonly used in real-time systems [26]. Our EDF algorithm satisfies the spatio-temporal constraints in the same way as BNT (lines 6 and 7 in Algorithm 1), but instead of performing the search described in Section 3.1, it visits nodes with the earliest time windows first. Consequently, assuming that the nodes are already sorted by time window, its time complexity is:

$$O\left(|V| \cdot |L| \cdot |A| \log |A|\right) \quad (14)$$

Below, we detail our setup and discuss the results.

### 4.1 Setup

Let $\mathcal{N}$ and $\mathcal{U}$ denote the normal and uniform distribution, respectively. A test configuration consists of the following parameters:

[1] https://doi.org/10.5281/zenodo.4728012
[2] https://doi.org/10.5281/zenodo.4728003

- Since there are currently 150 identical London fire engines in operation, $|A| = 150$ for each problem. All agents have the same speed, but each may perform differently in different coalitions.
- $|V| = |A| \cdot k$, where $k \in \mathbb{N}^+$ and $k \leq 20$. Thus, problems have up to 3000 nodes.
- The demand of each node $v$ is defined from a record dated between 1 January 2009 and 31 December 2020 as follows: $\alpha_v = \min_{a \in A} \rho(a, l_a^0, l_v)$; $\gamma_v = \alpha_v + \kappa$, where $\kappa$ is the attendance time (in seconds) of the firefighters; $\beta_v \sim \mathcal{U}(\alpha_v, \gamma_v)$; $w_v \sim \mathcal{U}(\kappa/2, \kappa)$; $|L_v| = 1$, and $\phi_v = 1$.
- For each node-to-agent ratio $|V|/|A|$, the nodes of a problem are chosen in chronological order. That is, the first problem always starts with record 1, and if a problem stops at record $q$, then the following one will use records $q+1$ to $q+1+|V|$.
- The locations are latitude-longitude points, and the travel time $\rho(a, l_1, l_2)$ is given by the distance between locations $l_1$ and $l_2$ divided by the (fixed) speed of agent $a$.
- In addition to node locations, $L$ contains the locations of the 103 currently active London fire stations. Each agent starts at a fire station defined by the record of a node.
- For each node $v_i$, with $i \leq |V| - 1$, if $\alpha_{v_i} \leq \alpha_{v_{i+1}}$ and $\gamma_{v_i} < \gamma_{v_{i+1}}$, then $v_i \rightarrow v_{i+1}$ has probability $1/2$.
- We use the following coalition value distributions:
  (1) *Superadditive*: $u(C, v, l) = |C|$.
  (2) *Uniform*: $u(C, v, l) \sim \mathcal{U}(0, |C|)$.
  (3) *Normal*: $u(C, v, l) \sim \mathcal{N}(10 \cdot |C|, 0.01)$.
  (4) *Modified Uniform*: $u(C, v, l) \sim \mathcal{U}(0, 10 \cdot |C|)$, and each $u(C, v, l)$ is also increased by $r \sim \mathcal{U}(0, 50)$ with probability $1/5$.
  (5) *Modified Normal*: like Normal, except that each $u(C, v, l)$ is also increased by $r \sim \mathcal{U}(0, 50)$ with probability $1/5$.
  (6) *Agent-based*: each agent $a$ has a value $p_a \sim \mathcal{U}(0, 10)$ representing its individual performance and a value $p_a^C \sim \mathcal{U}(0, 2 \cdot p_a)$ representing its performance in coalition $C$. The value of a coalition is the sum of the values of its members: $u(C, v, l) = \sum_{a \in C} p_a^C$.
  (7) *Normally Distributed Coalition Structures* (NDCS): $u(C, v, l) \sim \mathcal{N}(|C|, \sqrt[4]{|C|})$
  (8) *Congested NDCS*: like NDCS, except that each $\omega = u(C, v, l)$ is also decreased by $r \sim \mathcal{U}(\omega/10, \omega)$ with probability $|C|/(|A|+1)$.

Distribution 1 is typically used to study coalition formation problems [22], distributions 2 – 7 are taken from [18], and Congested NDCS is defined by us. We extended NDCS because it does not favor solutions containing fewer coalitions [19]. The additional perturbation simulates situations where the more agents there are, the greater the likelihood of congestion and thus of reduced performance, as it can happen in large-scale robot swarms [8].

For each test configuration and algorithm, we solved 100 problems and measured the median and 95% confidence interval of solution score (Equation 9) and CPU time. Since we have 2 algorithms, 20 node-to-agent ratios, 8 coalition value distributions and 100 replicates, the total number of tests performed is 32000. We ensured consistency between the results of the algorithms as follows. Regarding Superadditive, Uniform, Normal, Modified Uniform, and Modified Normal, all coalition values were computed and stored in hash maps before running the tests. Because such distributions depend only on coalition sizes, this preprocessing took $O(|A|)$ time. Regarding Agent-based, NDCS and Congested NDCS, the hash maps of coalition values were lazy-initialized.

### 4.2 Results

Figure 1 reports the results of our tests. The solution score tends to increase because it is an absolute metric, hence the bigger the problem, the more nodes there are that can be visited within their soft latest times.

Let $\eta = score(x_{BNT})/score(x_{EDF})$ be the *performance improvement* of BNT over EDF. For each coalition value distribution, the median value of $\eta$ over all node-to-agent ratios is greater than 1. We have $\eta \approx 1.53$ with Superadditive, $\eta \approx 1.29$ with Uniform and $\eta \approx 5.35$ with Normal (Figure 1a-c). All three distributions give more value to larger coalitions. More precisely, Superadditive does so by definition, while Uniform and Normal tend to generate solution spaces with small numbers of coalitions [19]. This motivates the use of BNT in real-time domains such as disaster response, where larger coalitions are typically more effective. The high value of $\eta$ with Normal is due to its mean, which has a constant factor of 10. We have approximately $3.5 \leq \eta \leq 5.41$ with Modified Uniform, Modified Normal and Agent-based (Figure 1d-f). This is because they are variants of Uniform and Normal, whose perturbations amplify their tendencies. NDCS and Congested NDCS reach approximately $1.43 \leq \eta \leq 1.5$ (Figure 1g, h), with the latter having the lowest values because it is a more penalizing variant. The reason why $\eta$ is lower with such distributions is that their solution spaces are much more difficult to prune than those of Normal, Uniform and their variants [19, Section 5.2]. Overall, we have $\eta \approx 2.52 \pm [0.73, 1.98]$.

Regarding CPU times[3], we recorded a median of $30 \pm [15, 26]$ ms for EDF and $15 \pm [24, 14.63]$ seconds for BNT. This was expected, since BNT is asymptotically slower than EDF (Equations 13 and 14).

To summarize, despite being up to 2 orders of magnitude slower than EDF, BNT can still solve problems with 150 agents and up to 3000 nodes within seconds, obtaining solution scores up to 3.25 times higher, both with distributions suited to disaster response (Figure 1a-f) and with difficult-to-prune distributions (Figure 1g, h).

## 5 CONCLUSIONS

We presented the MARSC problem, the first generalization of the TOPTW with coalition formation, precedences, multiple possible locations per node, workloads, and both soft and hard deadlines. We gave a binary integer program and designed BNT, the first anytime and scalable MARSC algorithm. Using a large-scale test framework and real-world data provided by the London Fire Brigade, we have shown that BNT can be used effectively in real-time domains such as disaster response.

Future work aims at defining a distributed MARSC algorithm with provable bounds on the solution quality, and creating the first anytime and exact algorithm to solve both the MARSC and the TOPTW [27]. Moreover, we want to extend our test framework by considering scenarios where nodes have heterogeneous profits and multiple possible locations, agents have different speeds, and coalition values also depend on nodes and locations. Two other limitations to be removed are as follows. First, we assume that the problem does not change over time and is known a priori. Second, we do not consider the impact of cross-schedule dependencies [15]. For instance, if the location chosen for node $v$ can no longer be accessed, or some agent cannot visit $v$ anymore due to unexpected failures, our model does not cope with the consequences for the nodes that are scheduled after $v$.

## ACKNOWLEDGEMENTS

This research is supported by UKRI and AXA Research Fund. Luca Capezzuto acknowledges the use of the IRIDIS High Performance Computing facility at the University of Southampton.

## A PROOF OF THEOREM 2.1

Based on [27, Section 3.3], we first describe the TOPTW using our terminology (Section 2.1), then we show that it is a special case of the MARSC. Consequently, this appendix also provides a novel mathematical programming formulation of the TOPTW.

Proof. The TOPTW considers a finite set of nodes, each with a non-negative profit. Each node has exactly one location and no soft latest time: $|L_v| = 1 \wedge \beta_v = \gamma_v, \forall v \in V$. When a node is visited within its time window, its profit is added to the total profit. Between each couple of nodes there is a fixed travel time. There are an initial node and a final node, both with zero profit, an earliest time of 0 and a soft latest time of $t_{max}$. Each route must begin at the initial node and end at the final node. The objective is to determine $n$ routes, one for each agent, that maximize the total profit.

The travel time between nodes $v_i$ and $v_j$ also includes the *service time* at $v_j$, that is, the time taken by any coalition to visit $v_j$ [27, Section 2.2]. Hence, we set $w_v = 1, \forall v \in V$, to ensure that each node is visited for at most 1 unit of time. Since travel times depend only on node locations, we exclude $A$ from the domain of $\rho(\cdot)$. Coalitions are singleton, that is, each $C$ is such that $|C| = 1$. The coalition value function $u(\cdot)$ always returns 1.

Because each node has exactly one location and coalitions are singleton, we remove the subscript $l$ from the decision variables and simply use $a$ instead of $C$ to indicate the coalition consisting of agent $a$. Hence, $x_{v,t,a} = 1$ if agent $a$ visits node $v$ at time $t$, and 0 otherwise. Let $v_{start}$ and $v_{end}$ denote respectively the initial and final nodes. The structural constraints (Equation 2) become:

$$\forall v \in V \setminus \{v_{start}, v_{end}\}, \forall t \in [\alpha_v, \beta_v], \sum_{a \in A} x_{v,t,a} \leq 1 \quad (15)$$

Since both coalition values and workloads are unitary, and each node does not have multiple locations, Equation 15 already ensures that a node is visited within its time window by at most one agent. Hence, no temporal constraints (Section 2.3.2) are required.

Let $l_{start}$ denote the location of $v_{start}$ and $l_v$ denote the location of any other node $v$. The initial location of each agent is $l_{start}$ and the spatial constraints (Equations 5 and 6) become:

$$\forall v \in V, \forall a \in A, \text{ if } \lambda = \rho(l_{start}, l_v) \geq \alpha_v \text{ then}$$

$$\sum_{t \in [\alpha_v, \lambda]} x_{v,t,a} = 0 \quad (16)$$

---

[3]Based on an Intel Xeon E5-2670 processor (octa-core 2.6 GHz with Hyper-Threading).

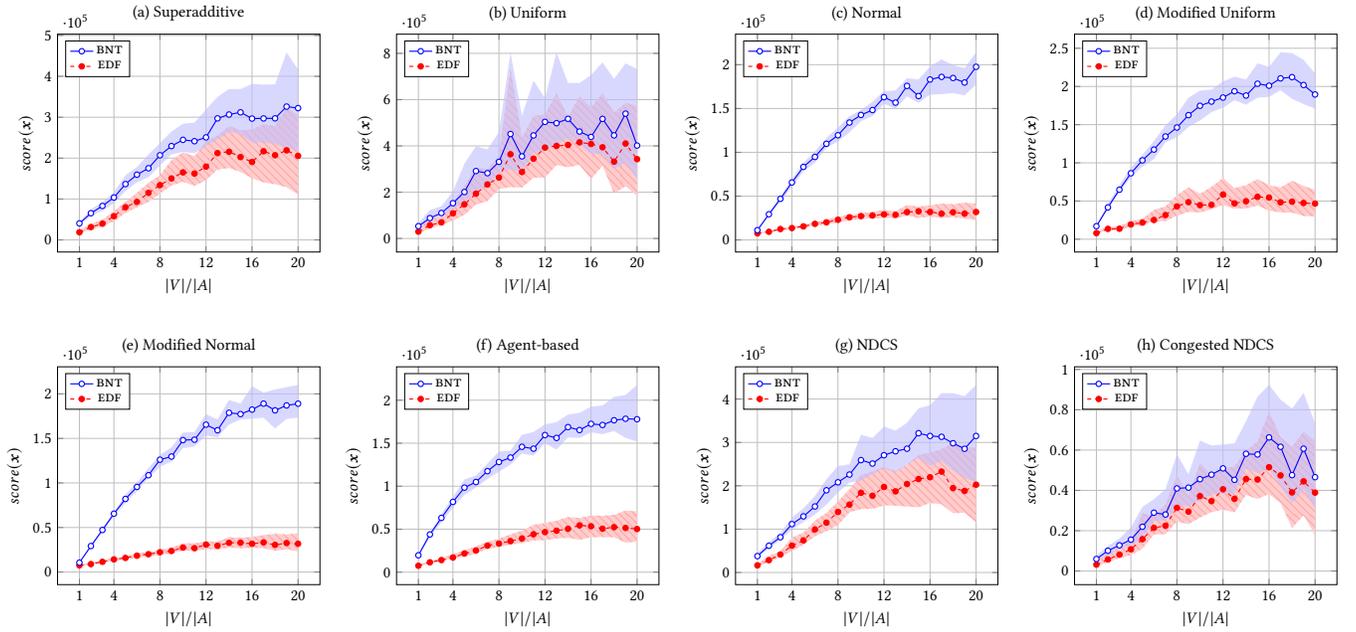

Figure 1: Performance of EDF and BNT in our tests. Each subfigure contains tests performed with the coalition value distribution in the title. Each point is the median and 95% confidence interval over 100 problems. The X-axis indicates node-to-agent ratios, while the Y-axis reports solution scores (Equation 9). The higher the scores, the better the solutions.

$$\forall v_1, v_2 \in V, \forall a \in A,$$
$$\forall t_1 \in [\alpha_1, \beta_1], \forall t_2 \in [\alpha_2, \beta_2] : t_1 + \rho(l_{v_1}, l_{v_2}) \geq t_2, \quad (17)$$
$$x_{v_1, t_1, a} + x_{v_2, t_2, a} \leq 1$$

To ensure that each route starts at $v_{start}$ and ends at $v_{end}$, we define the set of precedences as follows:

$$Prec = \{(v_{start}, v), (v, v_{end}) : v \in V \setminus \{v_{start}, v_{end}\}\}$$

The ordering constraints (Equation 7) become:

$$\forall v_1 \in V, \text{ if } \exists v_2 \in V : v_1 \rightarrow v_2 \wedge \alpha_2 \leq \beta_1 \text{ then}$$
$$\forall a \in A, \sum_{t \in [\alpha_2, \beta_1]} x_{v_2, t, a} = 0 \quad (18)$$

The absence of soft latest times means that there can be no time penalties: $\psi_{v,t} = 1, \forall v \in V, \forall t \leq t_{max}$ (Equation 8). Hence, the objective function (Equation 10) is simplified as follows:

$$\arg\max_{\mathbf{x}} \sum_{v \in V \setminus \{v_{start}, v_{end}\}} \sum_{t \in [\alpha_v, \beta_v]} \sum_{a \in A} \phi_v \cdot x_{v, t, a} \quad (19)$$
$$\text{subject to Equations 15 – 18}$$

□

## REFERENCES

[1] Alexander, D. E. *Principles of Emergency Planning and Management.* Oxford University Press, 2002.
[2] Burkard, R., Dell'Amico, M., and Martello, S. *Assignment Problems.* SIAM, 2012.
[3] Capezzuto, L., Tarapore, D., and Ramchurn, S. D. Anytime and efficient multi-agent coordination for disaster response. *SN Comput. Sci. 2*, 165 (2021).
[4] Cormen, T. H., Leiserson, C. E., Rivest, R. L., and Stein, C. *Introduction to Algorithms*, 3rd ed. MIT press, 2009.
[5] Gerkey, B. P., and Matarić, M. J. A formal analysis and taxonomy of task allocation in multi-robot systems. *Int. J. Rob. Res. 23*, 9 (2004), 939–954.
[6] Godoy, J., and Gini, M. Task allocation for spatially and temporally distributed tasks. In *IAS* (2013), Springer Berlin Heidelberg, pp. 603–612.
[7] Griva, I., Nash, S. G., and Sofer, A. *Linear and Nonlinear Optimization*, second ed. Society for Industrial and Applied Mathematics, 2009.
[8] Guerrero, J., Oliver, G., and Valero, O. Multi-robot coalitions formation with deadlines: Complexity analysis and solutions. *PloS one 12*, 1 (2017).
[9] Kitano, H., and Tadokoro, S. Robocup rescue: A grand challenge for multiagent and intelligent systems. *AI Magazine 22*, 1 (2001), 39–39.
[10] Korsah, G. A. *Exploring Bounded Optimal Coordination for Heterogeneous Teams with Cross-Schedule Dependencies.* PhD thesis, Carnegie Mellon University, 2011.
[11] Korsah, G. A., Stentz, A., and Dias, M. B. A comprehensive taxonomy for multi-robot task allocation. *Int. J. Rob. Res. 32*, 12 (2013), 1495–1512.
[12] London Datastore. LFB incident records, 2021. https://data.london.gov.uk/dataset/london-fire-brigade-incident-records.
[13] London Datastore. LFB mobilisation records, 2021. https://data.london.gov.uk/dataset/london-fire-brigade-mobilisation-records.
[14] Murphy, R. R. *Disaster Robotics.* MIT press, 2014.
[15] Nunes, E., Manner, M., Mitiche, H., and Gini, M. A taxonomy for task allocation problems with temporal and ordering constraints. *Rob. and Aut. Sys. 90* (2017), 55–70.
[16] Nunes, E., McIntire, M., and Gini, M. Decentralized multi-robot allocation of tasks with temporal and precedence constraints. *Adv. Rob. 31*, 22 (2017), 1193–1207.
[17] Öncan, T. A survey of the generalized assignment problem and its applications. *INFORMS 45*, 3 (2007), 123–141.
[18] Rahwan, T., Michalak, T., and Jennings, N. A hybrid algorithm for coalition structure generation. In *AAAI* (2012), vol. 26.
[19] Rahwan, T., Ramchurn, S. D., Jennings, N. R., and Giovannucci, A. An anytime algorithm for optimal coalition structure generation. *JAIR 34* (2009), 521–567.
[20] Ramchurn, S. D., Polukarov, M., Farinelli, A., Truong, C., and Jennings, N. R. Coalition formation with spatial and temporal constraints. In *AAMAS* (2010), pp. 1181–1188.
[21] Rizk, Y., Awad, M., and Tunstel, E. W. Cooperative heterogeneous multi-robot systems: A survey. *ACM Comp. Sur. 52*, 2 (Apr. 2019).
[22] Sandholm, T. W., Larson, K., Andersson, M., Shehory, O., and Tohmé, F. Coalition structure generation with worst case guarantees. *Artificial Intelligence 111*, 1-2 (1999), 209–238.
[23] Scerri, P., Farinelli, A., Okamoto, S., and Tambe, M. Allocating tasks in extreme teams. In *AAMAS* (2005), pp. 727–734.

[24] Shehory, O., and Kraus, S. Methods for task allocation via agent coalition formation. *AI 101*, 1-2 (1998), 165–200.
[25] Shoham, Y., and Leyton-Brown, K. *Multiagent Systems: Algorithmic, Game-Theoretic, and Logical Foundations*. Cambridge University Press, 2008.
[26] Stankovic, J. A., Spuri, M., Ramamritham, K., and Buttazzo, G. C. *Deadline scheduling for real-time systems: EDF and related algorithms*, vol. 460. Springer Science & Business Media, 2013. Reprint of the original 1998 edition.
[27] Vansteenwegen, P., and Gunawan, A. *Orienteering Problems: Models and Algorithms for Vehicle Routing Problems with Profits*. Springer Nature, Switzerland, 2019.
[28] Vig, L., and Adams, J. A. Multi-robot coalition formation. *IEEE Trans. on Rob. 22*, 4 (2006), 637–649.
[29] Wolsey, L. A. *Integer Programming*, second ed. John Wiley & Sons, 2020.
[30] Zlot, R. M. *An Auction-Based Approach to Complex Task Allocation for Multirobot Teams*. PhD thesis, Carnegie Mellon University, 2006.